\documentclass[twocolumn,twoside,slac_two]{revtex4}
\usepackage{graphicx}
\usepackage{fancyhdr}
\newcommand {\GeV}       {\mbox{GeV}}
\newcommand {\TeV}       {\mbox{TeV}}

\newcommand {\pT}        {\mbox{$p_T$\ }}
\newcommand {\pTZ}       {\mbox{$p_T^Z$\ }}
\newcommand {\pTjet}     {\mbox{$p_T^{jet}$\ }}
\newcommand {\ptjet}     {\mbox{$p_T^{jet}$\ }}
\newcommand {\yZ}        {\mbox{$y^{Z}$}\ }
\newcommand {\yjet}      {\mbox{$y^{jet}$}\ }
\newcommand{\Dzero}      {\mbox{D{\O}\ }}

\newcommand{\Zgammas}    {\mbox{$Z\slash \gamma^{*}\ $}}
\newcommand{\Zmumu}      {\mbox{$Z\slash \gamma^{*} \rightarrow \mu\mu\ $}}
\newcommand{\Zee}        {\mbox{$Z\slash \gamma^{*} \rightarrow ee\ $}}

\begin{document}

\title{Measurements of Differential {\bf \Zgammas+jet+X} Cross Sections with the \Dzero\ Detector}


\author{Sabine Lammers (on behalf of of the \Dzero\ collaboration)}
\affiliation{Department of Physics, Indiana University, Bloomington, IN 47405, USA}

\begin{abstract}
We present measurements of differential cross sections in inclusive $Z\slash \gamma^*$ plus jet production in a data sample of 1 $fb^{-1}$ collected with the \Dzero\ detector in proton antiproton collisions at 
$\sqrt{s}$ = 1.96 $\TeV$.  Measured variables include the \Zgammas transverse momentum (\pTZ) and rapidity (\yZ), the leading jet transverse momentum (\pTjet) and rapidity (\yjet), as well as various
angles of the Z+jet system.   We compare the results to different Monte Carlo event generators and 
next-to-leading order perturbative QCD (NLO pQCD) predictions, with non-perturbative corrections applied.
\end{abstract}

\maketitle

\thispagestyle{fancy}


%%%%%%%%%%%%%%%%%%%%%%%%%%%%%%%%%%
\section{Introduction}
To make discoveries at the Tevatron and the LHC, background processes will need to be measured and simulated with a level of accuracy that will be comparable to the signficance of those new physics signals.  There are several programs on the market that can simulate hadronic interactions at next-to-leading order (NLO) accuracy, but the processes included in these programs are limited.  Matrix element plus parton shower (MEPS) programs simulate a more comprehensive set of processes, typically at leading-log (LL) or leading order (LO), and rely on models to simulate emissions and fragmentation associated with higher order processes.  These programs have been employed regularly for background simulation at the Tevatron in recent years, notably in the Higgs searches \cite{Abazov:2007hk} and the discovery of the production of single top quarks \cite{Abazov:2009ii}.  

Measurements of \Zgammas + jets processes are valuable for two principle reasons.  
\Zgammas production provides a hard scale, which, along with associated jet production, is an ideal environment to test perturbative QCD (pQCD).  The leptonic decay of the \Zgammas provides a clean signal for reconstruction of the events, and small background contamination.  The test of pQCD is made by comparing the measurements to NLO pQCD predictions. \Zgammas + jets also makes up a major background of many new physics searches at both the Tevatron and LHC.  Therefore, these data measurements unfolded to the particle level are useful for tuning LO simulation programs which are heavily relied upon to  model background processes.  

The Tevatron measurements presented here of \Zgammas + jets differential cross sections are compared to predictions by NLO pQCD in MCFM \cite{Campbell:2002tg}, MEPS programs {\sc ALPGEN} \cite{Mangano:2002ea} and {\sc SHERPA} \cite{Gleisberg:2008ta}, and PS programs {\sc HERWIG} \cite{Corcella:2000bw} and {\sc PYTHIA} \cite{Sjostrand:2000wi}.  The measurements have either been published \cite{Abazov:2008ez} \cite{Abazov:2009av} or have been submitted for publication \cite{Abazov:2009pp} at the time these proceedings were written.  ALPGEN employs the MLM algorithm to ensure jets originating from the matrix element and the parton shower are not double counted.  SHERPA is a CKKW-inspired model which uses a reweighting of the matrix elements to achieve the same appropriate jet configurations.  A detailed description of these programs can be found in \cite{Alwall}.

%%%%%%%%%%%%%%%%%%%%%%%%%%%%%%%%%%
\section{Data Selection}

The measurements are made with the \Dzero detector, which is described in detail elsewhere \cite{Abazov:2005pn}.  The analyses were performed in the \Zmumu and \Zee decay channels.  \Zmumu focused on the Z+1 jet inclusive events, and differential cross sections were made in a variety of variables, including many angular  variables involving the decay objects.  The \Zee analysis focused on differential cross section measurements as a function of \ptjet, in the 1, 2 and 3 jet inclusive samples.  In both channels, the data are corrected to particle level to eliminate the effects arising from detector resolution and efficiency.  In order to keep these corrections small, a limited particle level phase space for the measurements was chosen that corresponds closely to the detector level selection cuts.  Some effort was made to keep the particle level phase space in the \Zmumu and \Zee channels close, but some differences exist due to the different nature and detector manifestations of electrons and muons.  At particle level, the \Zgammas is reconstructed using the two highest \pT leptons, and the jets are reconstructed using the \Dzero RunII Midpoint Cone algorithm with a cone radius R=0.5.  In the \Zmumu analysis, the particle level phase space is defined with the following restrictions:

\begin{itemize}
\item $ 65 < M_{\mu\mu} < 115$ GeV
\item $\ptjet > 20$ GeV
\item $|y^{jet}| < 2.8$
\item $|y^{\mu}| < 1.7$ 
\end{itemize}

In the \Zee analysis, the particle level phase space is defined with the following restrictions:

\begin{itemize}
\item $ 65 < M_{ee} < 115$ GeV
\item $\ptjet > 20$ GeV
\item $|y^{jet}| < 2.5$
\end{itemize}

%%%%%%%%%%%%%%%%%%%%%%%%%%%%%%%%%%
\section{Theoretical Predictions}

Due to the evolving nature of theoretical predictions, the programs used to compare to the data are not consistent in all distributions.  This is due to the fact that the data are published at different times, and effort is made to always compare to the most up-to-date theoretical predictions.  Next-to-leading order perturbative QCD (NLO pQCD) predictions are calculated using the program MCFM; parton-to-hadron corrections are applied to the MCFM predictions, as calculated by \textsc{PYTHIA}.  

The meausurements of \Zmumu + jets are compared to different predictions than those compared in the \Zee + jets measurements.  The programs and their version numbers are summarized in Table \ref{example_table}. 

\begin{table}[h]
\begin{center}
\caption{Theoretical predictions and their versions for \Zmumu and \Zee + jets data comparisons.}
\vspace{2mm}
\begin{tabular}{|l|c|c|}
\hline \textbf{Program} & \textbf{\Zmumu} & \textbf{\Zee} \\
\hline {\sc MCFM} & 5.4 & 5.3 \\
\hline {\sc ALPGEN} & 2.13 & 2.13 \\
\hline {\sc SHERPA} & 1.1.3 & 1.1.1 \\
\hline {\sc PYTHIA} & 6.420 & 6.416\footnote{PYTHIA 6.325 is used for interfacing ALPGEN with parton showers} \\
\hline {\sc HERWIG} & 6.510 & 6.510 \\
\hline
\end{tabular}
\label{example_table}
\end{center}
\end{table}

%%%%%%%%%%%%%%%%%%%%%%%%%%%%%%%%%%
\section{Results}

Particle level differential cross sections as functions of a variety of variables are shown in Figures 1-13 for \Zgammas + jets events.  Figures 1-4 and 8-13 illustrate measurements that were made in the \Zmumu decay channel, while Figures 5-7 show measurements that were made in the \Zee decay channel.  
Figures 8, 10 and 12 consider \Zgammas + jets events in the restricted range \Zgammas $>$ 25 GeV, while Figures 9, 11 and 13 restrict this range further to \Zgammas $>$ 45 GeV.  These restrictions are made to reduce the dependence on the underlying event and multiple interactions.   All \Zmumu results are made with \Zgammas + 1 jet inclusive samples.  The \Zee results examine the \Zgammas + 1, 2 and 3 jet inclusive samples.  A good description of the data by NLO pQCD is found in all distributions, taking into account the experimental and theoretical uncertainties.  {\sc SHERPA} generally gives the best and most consistent description of the angular distributions, although the normalization of the data is not reproduced.  The description of the data by the LO MC models is mixed.  All LO MC models suffer from large scale uncertainties which impair their ability to make precise predictions.  

%%%%%%%%%%%%%%%%%%%%%%%%%%%%%%%%%%
\section{Conclusions}

Several differential cross sections of \Zgammas + jet + X events measured with the \Dzero detector have been presented.  The data are generally consistent with predictions from NLO pQCD, although some LO programs can reproduce the shape of the data better than NLO, due either to their inclusion of higher parton multiplicity matrix elements than can be currently included in a fixed order pQCD calculation, or an optimized tune of the Monte Carlo.  These data should be useful for continued tuning of these and other Monte Carlo programs.

\begin{acknowledgments}
The author would like to thank Gavin Hesketh for providing the plots with updated theoretical predictions and for general good humor.  
\end{acknowledgments}

\bigskip % extra skip inserted

%%%%%%%%%%%%%%%%%%%%%%%%%%%%%%%%%%

% figures should be put into the text as floats.
% Use the graphics or graphicx packages (distributed with LaTeX2e)
% and the \includegraphics macro defined in those packages.
% See the LaTeX Graphics Companion by Michel Goosens, Sebastian Rahtz,
% entire page. There is no need to do explicit centering.

% \begin{figure}
% \end{figure}

% Surround figure environment with turnpage environment for landscape
% figure
% \begin{figure}
% \end{figure}

\clearpage

\begin{figure}[p]
\centering
\includegraphics[width=80mm]{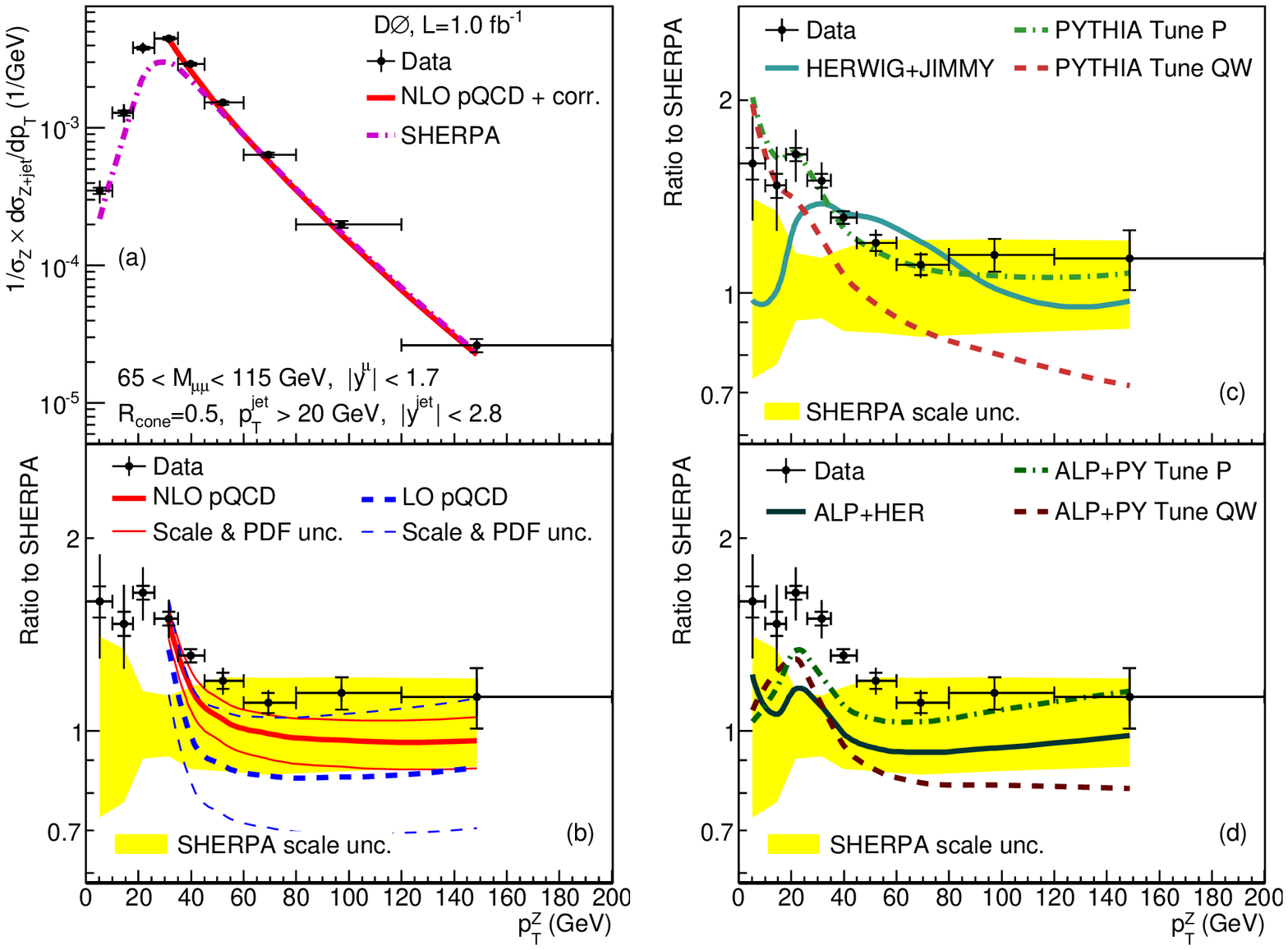}
\caption{The measured cross section in bins of leading \pTZ for \Zgammas + jet + X events, and  predictions from NLO pQCD and {\sc SHERPA} are shown in the upper left plot.  The ratio of data to {\sc SHERPA} are shown in the lower left plot.  The ratio of LO and NLO MCFM predictions to \textsc{SHERPA} and associated scale and PDF uncertainties are also shown in the lower left plot.  The ratio of \textsc{PYTHIA} and \textsc{HERWIG}+\textsc{JIMMY} to \textsc{SHERPA} are shown in the upper right plot.  The ratio of\textsc{ALPGEN}+\textsc{PYTHIA} and \textsc{ALPGEN}+\textsc{HERWIG} to \textsc{SHERPA} are shown in the lower right plot.  The ratio of data to \textsc{SHERPA} are included in all plots for comparison. }
\label{fig1}
\end{figure}

\begin{figure}[h]
\centering
\includegraphics[width=80mm]{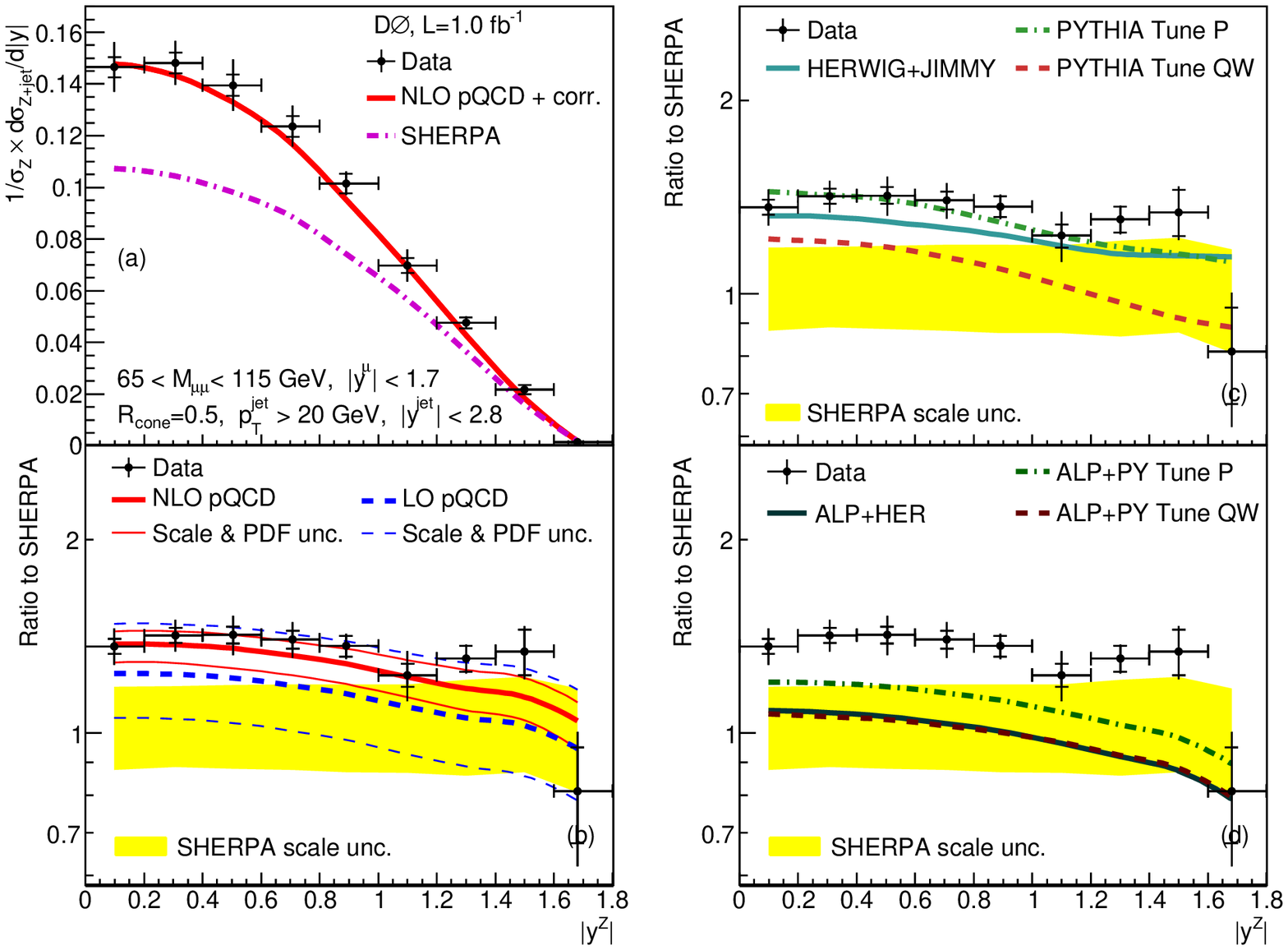}
\caption{The measured cross section in bins of leading \yZ for \Zgammas + jet + X events, and  predictions from NLO pQCD and {\sc SHERPA} are shown in the upper left plot.  The ratio of data to {\sc SHERPA} are shown in the lower left plot.  The ratio of LO and NLO MCFM predictions to \textsc{SHERPA} and associated scale and PDF uncertainties are also shown in the lower left plot.  The ratio of \textsc{PYTHIA} and \textsc{HERWIG}+\textsc{JIMMY} to \textsc{SHERPA} are shown in the upper right plot.  The ratio of\textsc{ALPGEN}+\textsc{PYTHIA} and \textsc{ALPGEN}+\textsc{HERWIG} to \textsc{SHERPA} are shown in the lower right plot.  The ratio of data to \textsc{SHERPA} are included in all plots for comparison. } 
\label{fig2}
\end{figure}

\begin{figure}[h]
\centering
\includegraphics[width=80mm]{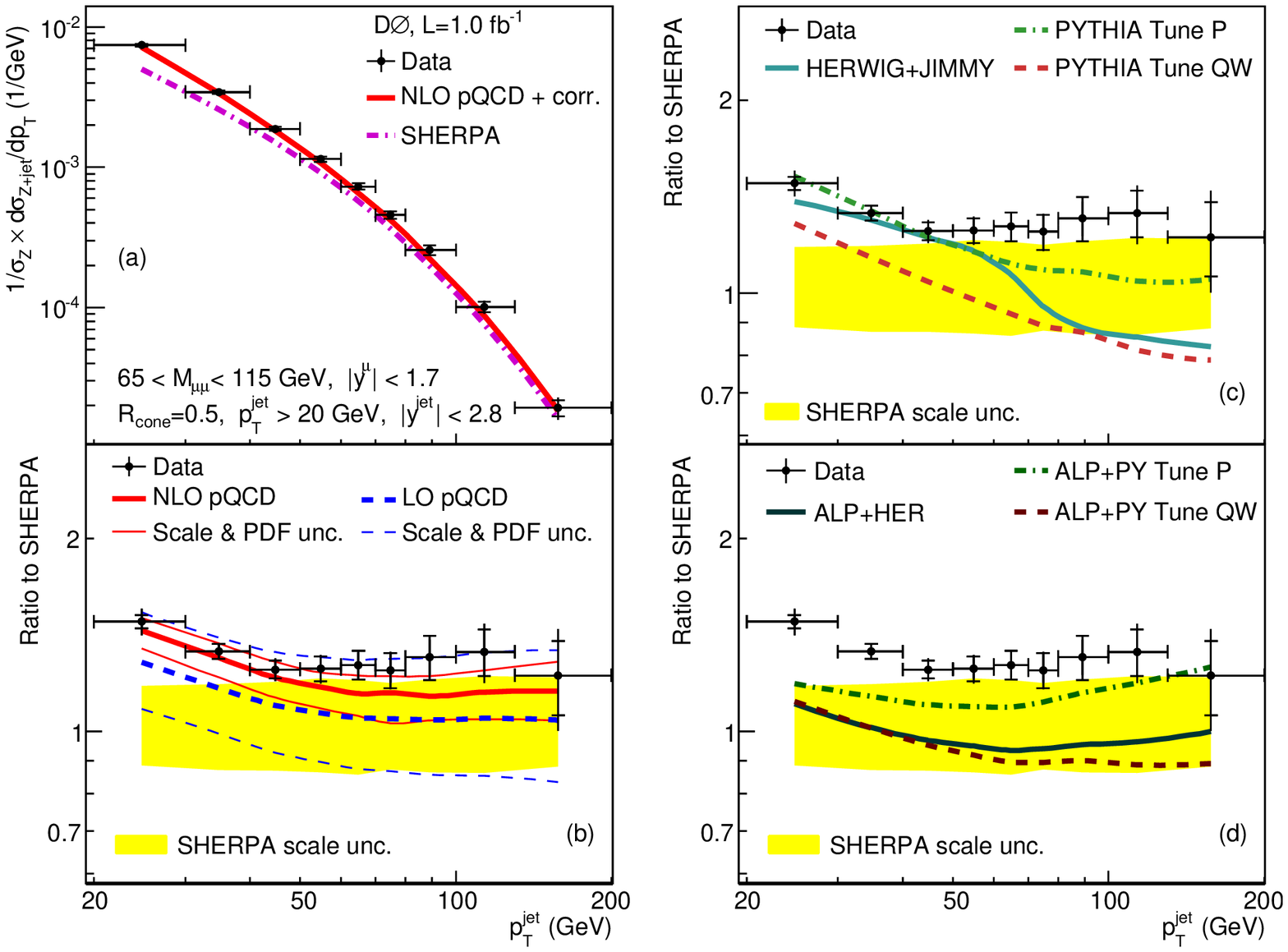}
\caption{The measured cross section in bins of leading \ptjet for \Zgammas + jet + X events, and  predictions from NLO pQCD and {\sc SHERPA} are shown in the upper left plot.  The ratio of data to {\sc SHERPA} are shown in the lower left plot.  The ratio of LO and NLO MCFM predictions to \textsc{SHERPA} and associated scale and PDF uncertainties are also shown in the lower left plot.  The ratio of \textsc{PYTHIA} and \textsc{HERWIG}+\textsc{JIMMY} to \textsc{SHERPA} are shown in the upper right plot.  The ratio of\textsc{ALPGEN}+\textsc{PYTHIA} and \textsc{ALPGEN}+\textsc{HERWIG} to \textsc{SHERPA} are shown in the lower right plot.  The ratio of data to \textsc{SHERPA} are included in all plots for comparison. } 
\label{fig3}
\end{figure}

\begin{figure}[h]
\centering
\includegraphics[width=80mm]{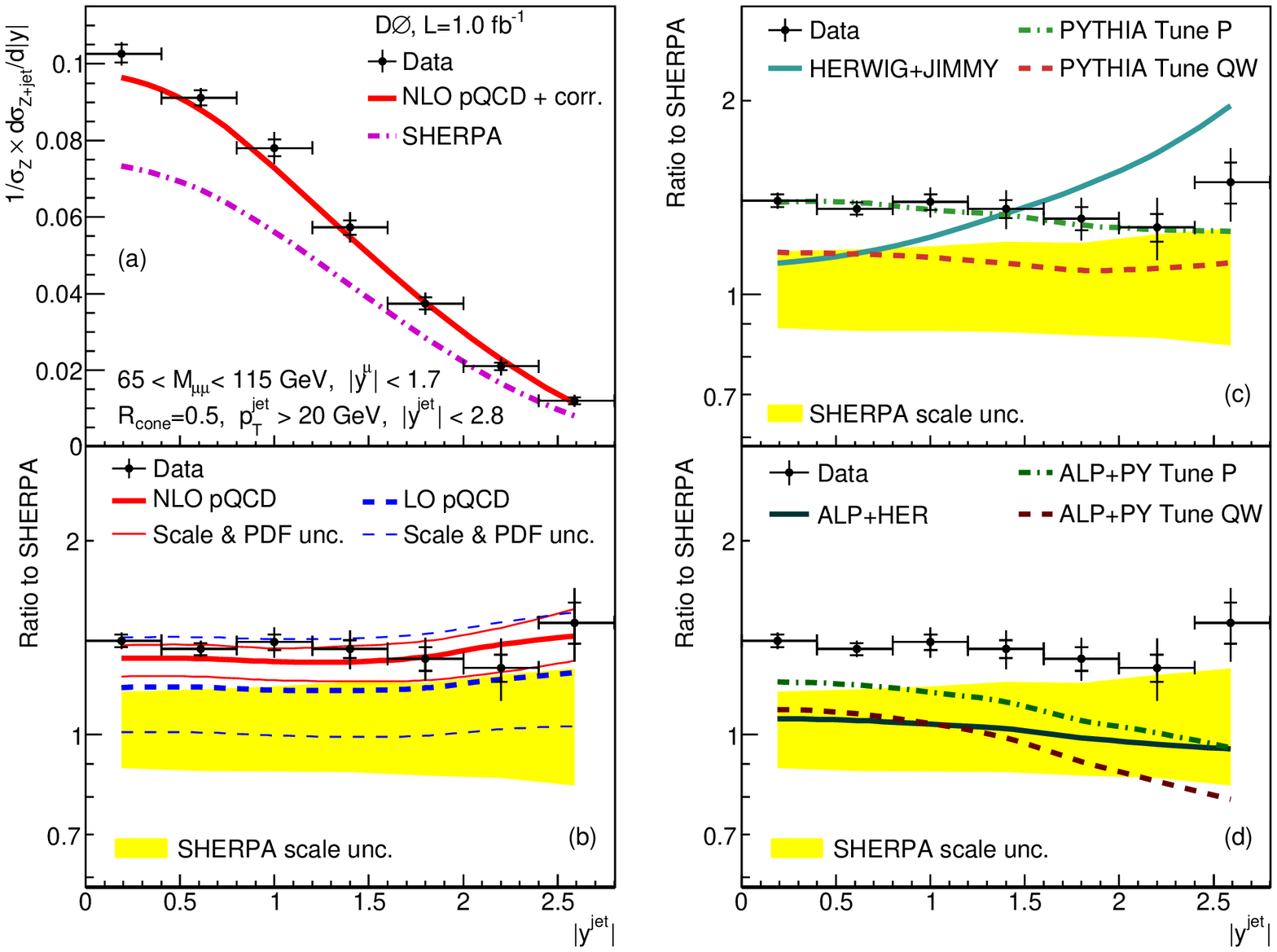}
\caption{The measured cross section in bins of leading \yjet for \Zgammas + jet + X events, and  predictions from NLO pQCD and {\sc SHERPA} are shown in the upper left plot.  The ratio of data to {\sc SHERPA} are shown in the lower left plot.  The ratio of LO and NLO MCFM predictions to \textsc{SHERPA} and associated scale and PDF uncertainties are also shown in the lower left plot.  The ratio of \textsc{PYTHIA} and \textsc{HERWIG}+\textsc{JIMMY} to \textsc{SHERPA} are shown in the upper right plot.  The ratio of\textsc{ALPGEN}+\textsc{PYTHIA} and \textsc{ALPGEN}+\textsc{HERWIG} to \textsc{SHERPA} are shown in the lower right plot.  The ratio of data to \textsc{SHERPA} are included in all plots for comparison. } 
\label{fig4}
\end{figure}

\clearpage

\begin{figure}[h]
\centering
\includegraphics[width=80mm]{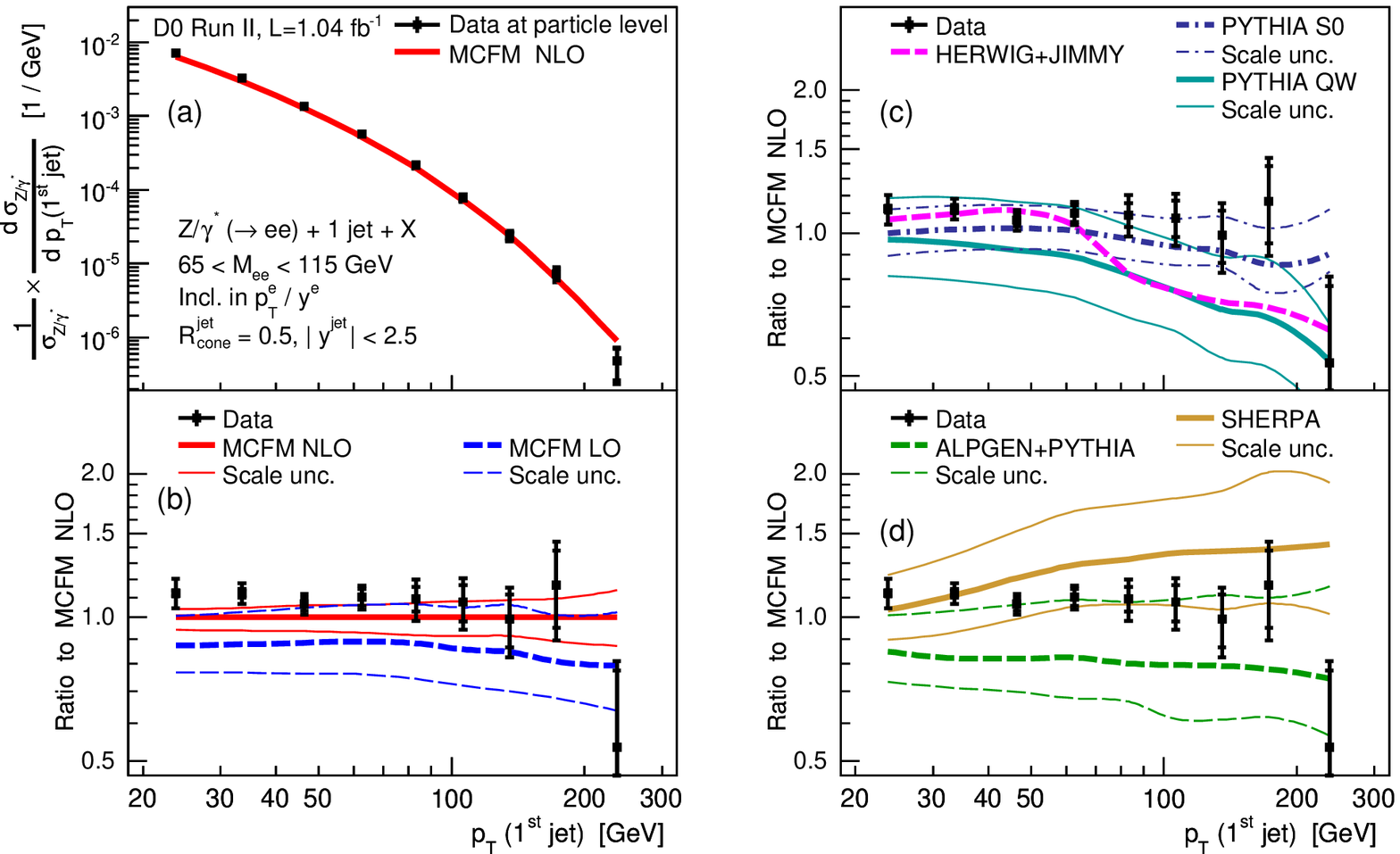}
\caption{The measured cross section for \Zgammas + 1 jet + X events in bins of leading \pTjet.  Predictions from NLO pQCD are compared to the data in the upper left plot. The ratio of data and several LO programs to MCFM NLO are shown in the other plots.  MCFM NLO predictions agree with the data.  The data can be described by a selection of the LO programs, although there is a lot of freedom in the predictions due to the choice of PDF, renormalization and factorization scale, tune and underlying event model.} 
\label{fig5}
\end{figure}

\begin{figure}[h]
\centering
\includegraphics[width=80mm]{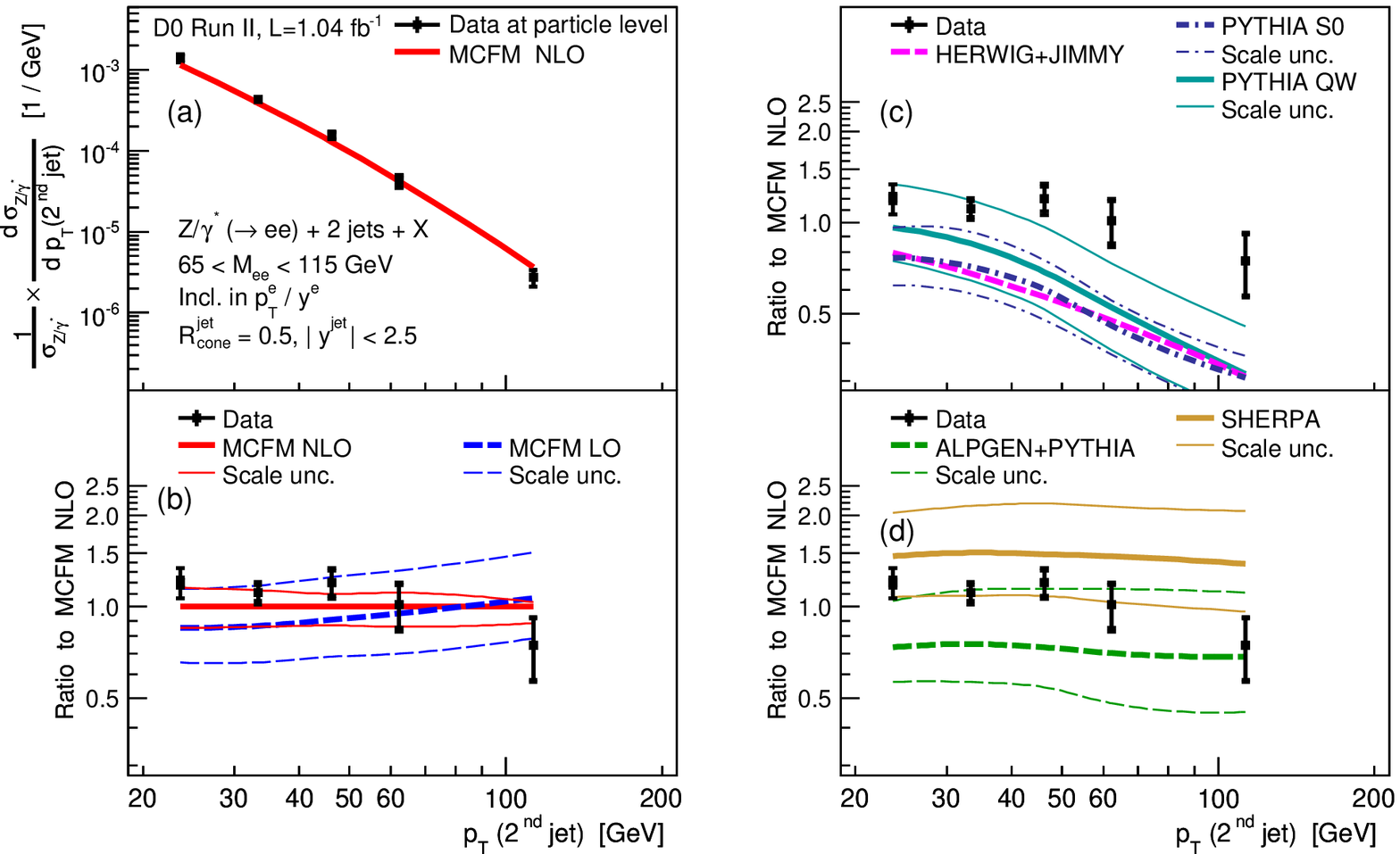}
\caption{The measured cross section for \Zgammas + 2 jet + X events in bins of the second leading \pTjet.  Predictions from NLO pQCD are compared to the data in the upper left plot. The ratio of data and several LO programs to MCFM NLO are shown in the other plots.  MCFM NLO predictions agree with the data.  The data can be described by a selection of the LO programs, although there is a lot of freedom in the predictions due to the choice of PDF, renormalization and factorization scale, tune and underlying event model.} 
\label{fig6}
\end{figure}

\begin{figure}[h]
\centering
\includegraphics[width=80mm]{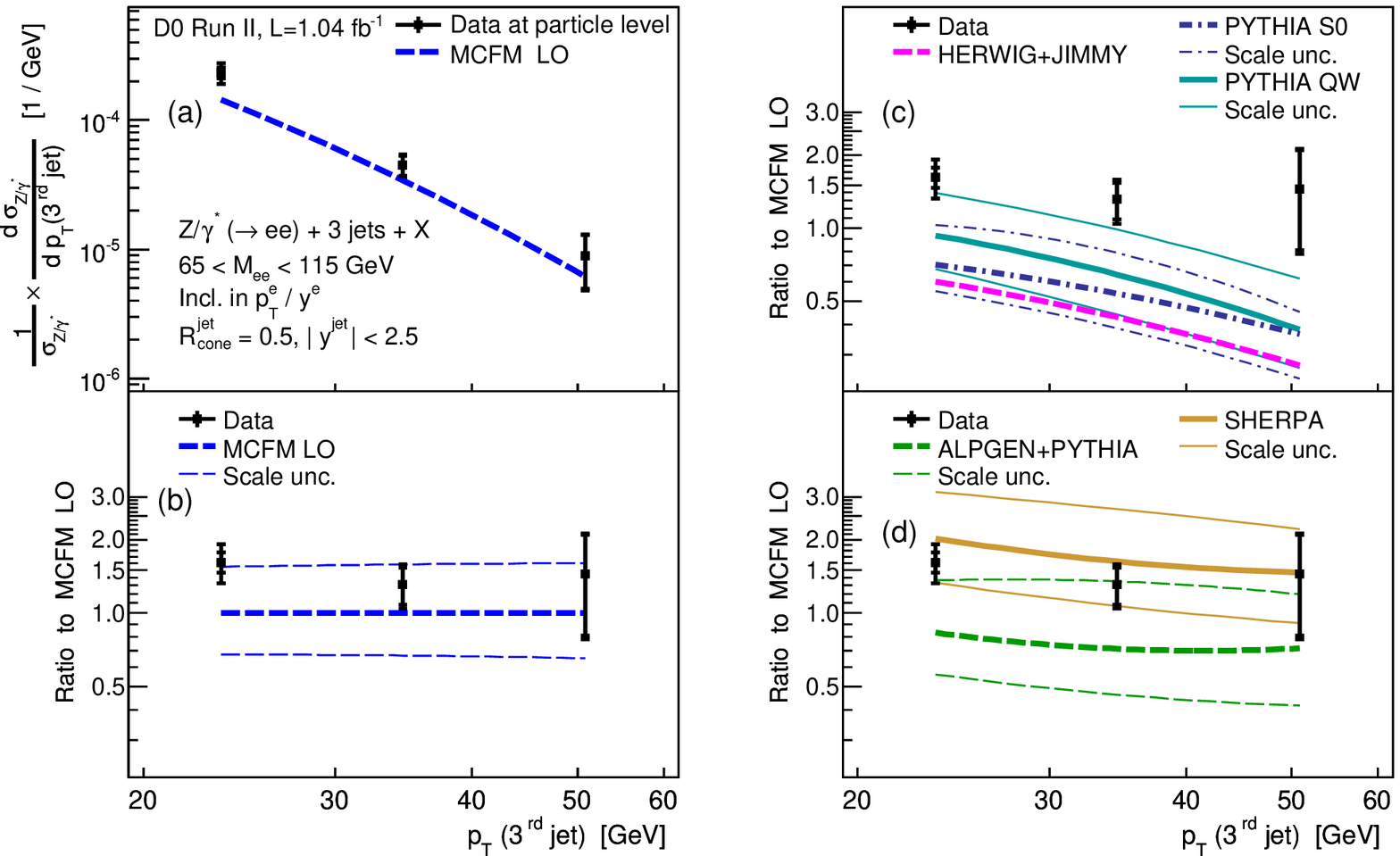}
\caption{The measured cross section for \Zgammas + 3 jet + X events in bins of the third leading \pTjet.  Predictions from NLO pQCD are compared to the data in the upper left plot. The ratio of data and several LO programs to MCFM NLO are shown in the other plots.  MCFM NLO predictions agree with the data.  The data can be described by a selection of the LO programs, although there is a lot of freedom in the predictions due to the choice of PDF, renormalization and factorization scale, tune and underlying event model.} 
\label{fig7}
\end{figure}

\clearpage

\begin{figure}[h]
\centering
\includegraphics[width=80mm]{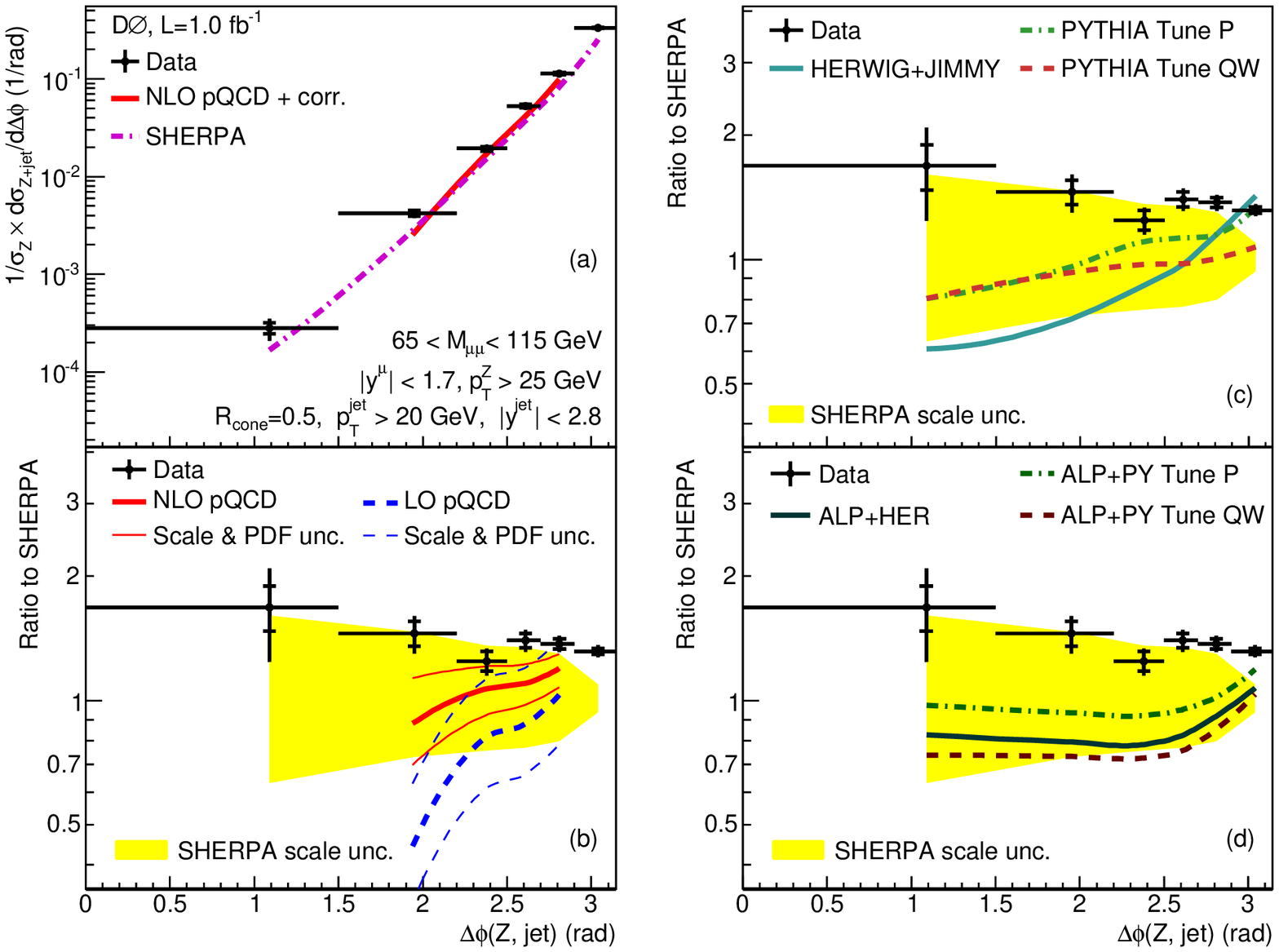}
\caption{The measured cross section in bins of $\Delta \Phi$ between the \Zgammas and leading jet for \Zgammas + jet + X events with \pTZ larger than 25 \GeV.  Predictions from NLO pQCD and \textsc{SHERPA} are compared to the data in the  upper left plot.  The ratio of data and predictions from NLO pQCD, \textsc{ALPGEN}, \textsc{HERWIG} and \textsc{PYTHIA} to the prediction from \textsc{SHERPA} are shown in the lower plots.  The shape of the distribution is best described by \textsc{SHERPA}, but there is a normalization disparity. } 
\label{fig8}
\end{figure}

\begin{figure}[h]
\centering
\includegraphics[width=80mm]{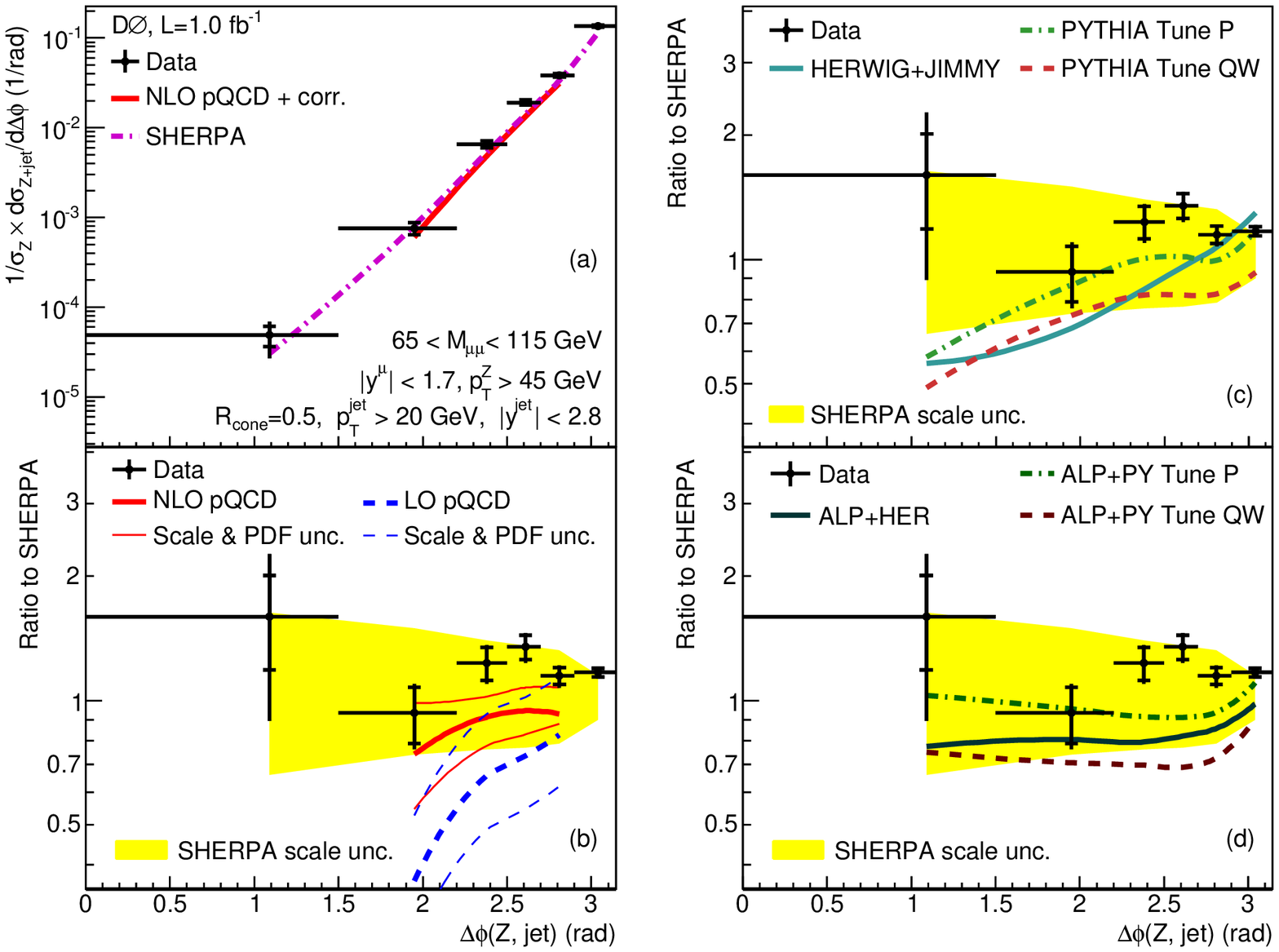}
\caption{The measured cross section in bins of $\Delta \Phi$ between the \Zgammas and leading jet for \Zgammas + jet + X events with \pTZ larger than 45 \GeV.  The restricted \pTZ range is chosen to isolate the measurement from biases due to the underlying event.  Predictions from NLO pQCD and \textsc{SHERPA} are compared to the data in the  upper left plot.  The ratio of data and predictions from NLO pQCD, \textsc{ALPGEN}, \textsc{HERWIG} and \textsc{PYTHIA} to the prediction from \textsc{SHERPA} are shown in the lower plots.  } 
\label{fig9}
\end{figure}

\begin{figure}[h]
\centering
\includegraphics[width=80mm]{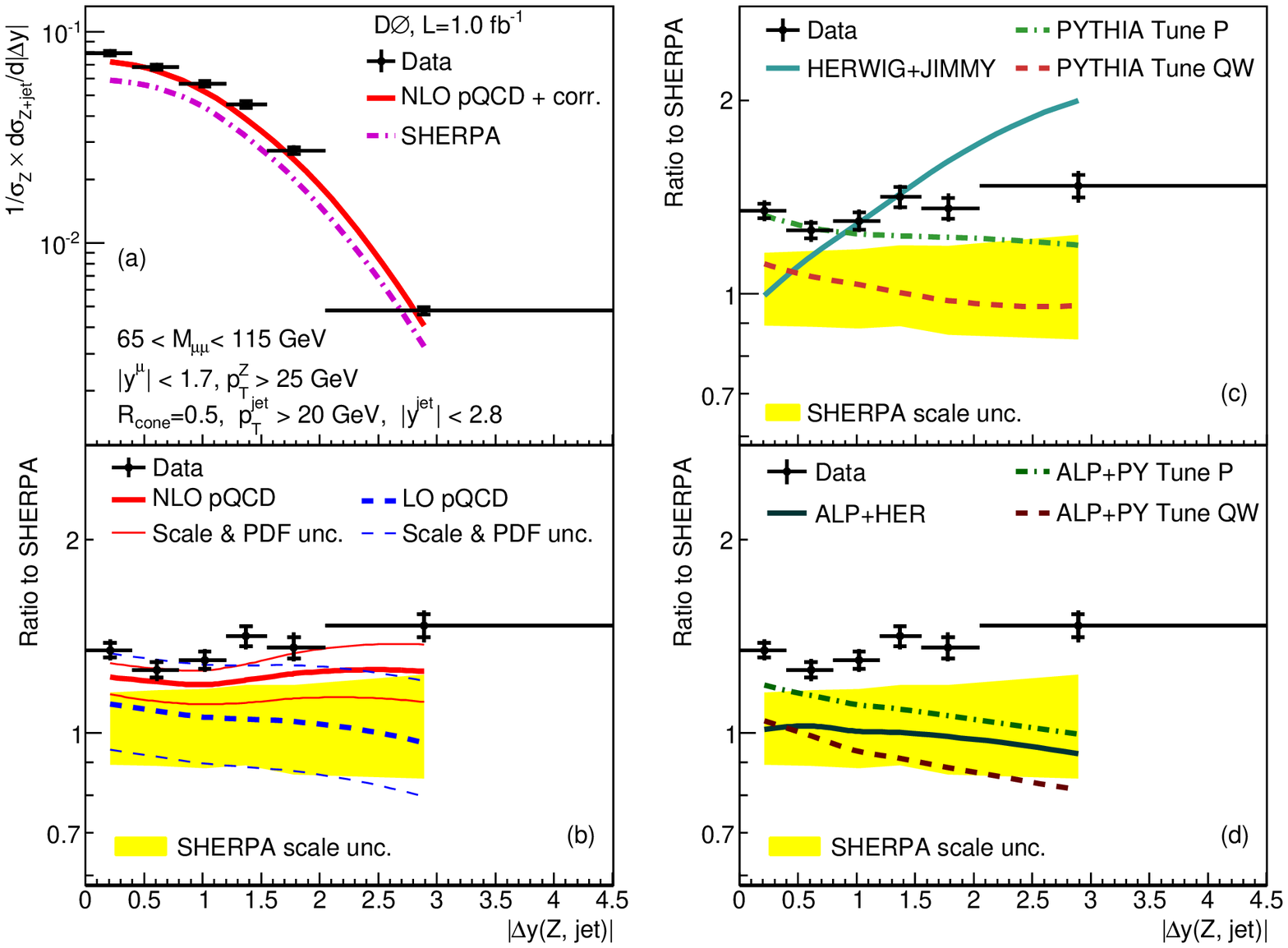}
\caption{The measured cross section in bins of $\Delta y$ between the \Zgammas and leading jet for \Zgammas + jet + X events with \pTZ larger than 25 \GeV.  Predictions from NLO pQCD and \textsc{SHERPA} are compared to the data in the  upper left plot.  The ratio of data and predictions from NLO pQCD, \textsc{ALPGEN}, \textsc{HERWIG} and \textsc{PYTHIA} to the prediction from \textsc{SHERPA} are shown in the lower plots. } 
\label{fig10}
\end{figure}

\begin{figure}[h]
\centering
\includegraphics[width=80mm]{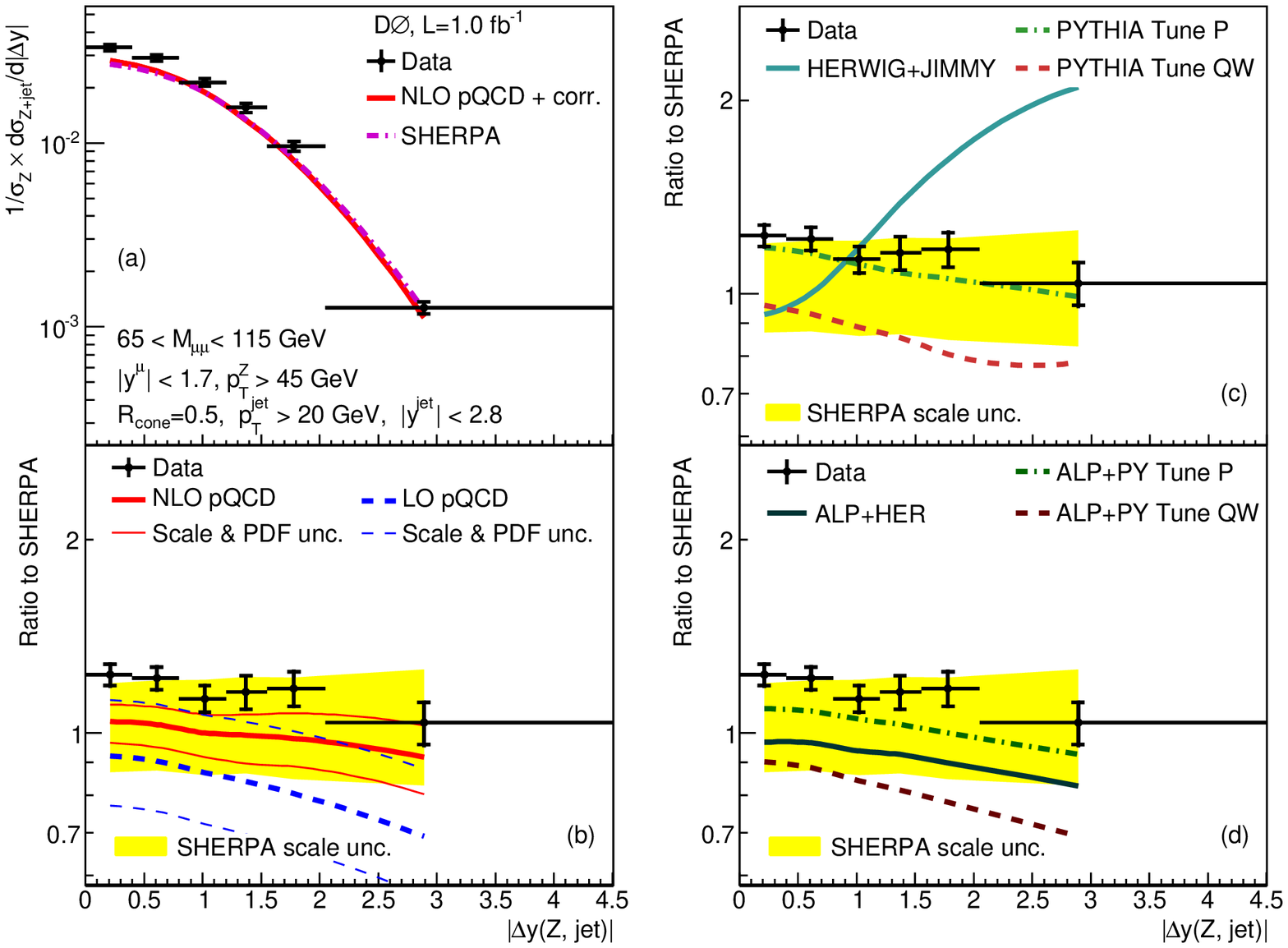}
\caption{The measured cross section in bins of $\Delta y$ between the \Zgammas and leading jet for \Zgammas + jet + X events with \pTZ larger than 45 \GeV.  The restricted \pTZ range is chosen to isolate the measurement from biases due to the underlying event.  Predictions from NLO pQCD and \textsc{SHERPA} are compared to the data in the  upper left plot.  The ratio of data and predictions from NLO pQCD, \textsc{ALPGEN}, \textsc{HERWIG} and \textsc{PYTHIA} to the prediction from \textsc{SHERPA} are shown in the lower plots. } 
\label{fig11}
\end{figure}

\clearpage

\begin{figure}[h]
\centering
\includegraphics[width=80mm]{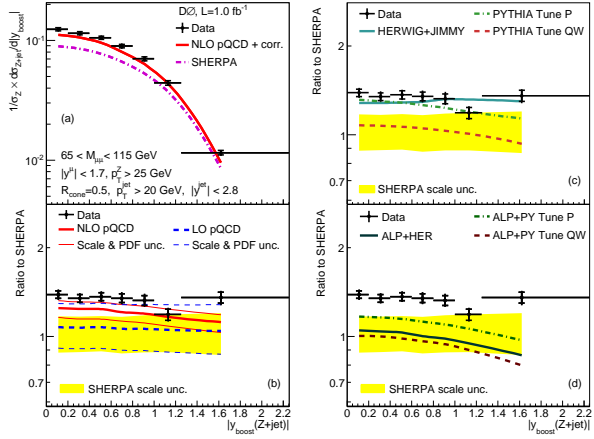}
\caption{The measured cross section in bins of $y_{boost}$ for \Zgammas + jet + X events with \pTZ larger than 25 \GeV, where $y_{boost}$ is the average rapidity of the reconstructed Z boson and the leading jet rapidity.  Predictions from NLO pQCD and \textsc{SHERPA} are compared to the data in the  upper left plot.  The ratio of data and predictions from NLO pQCD, \textsc{ALPGEN}, \textsc{HERWIG} and \textsc{PYTHIA} to the prediction from \textsc{SHERPA} are shown in the lower plots.  The shape of the distribution is best described by \textsc{SHERPA}, but there is a normalization disparity. } 
\label{fig12}
\end{figure}

\begin{figure}[h]
\centering
\includegraphics[width=80mm]{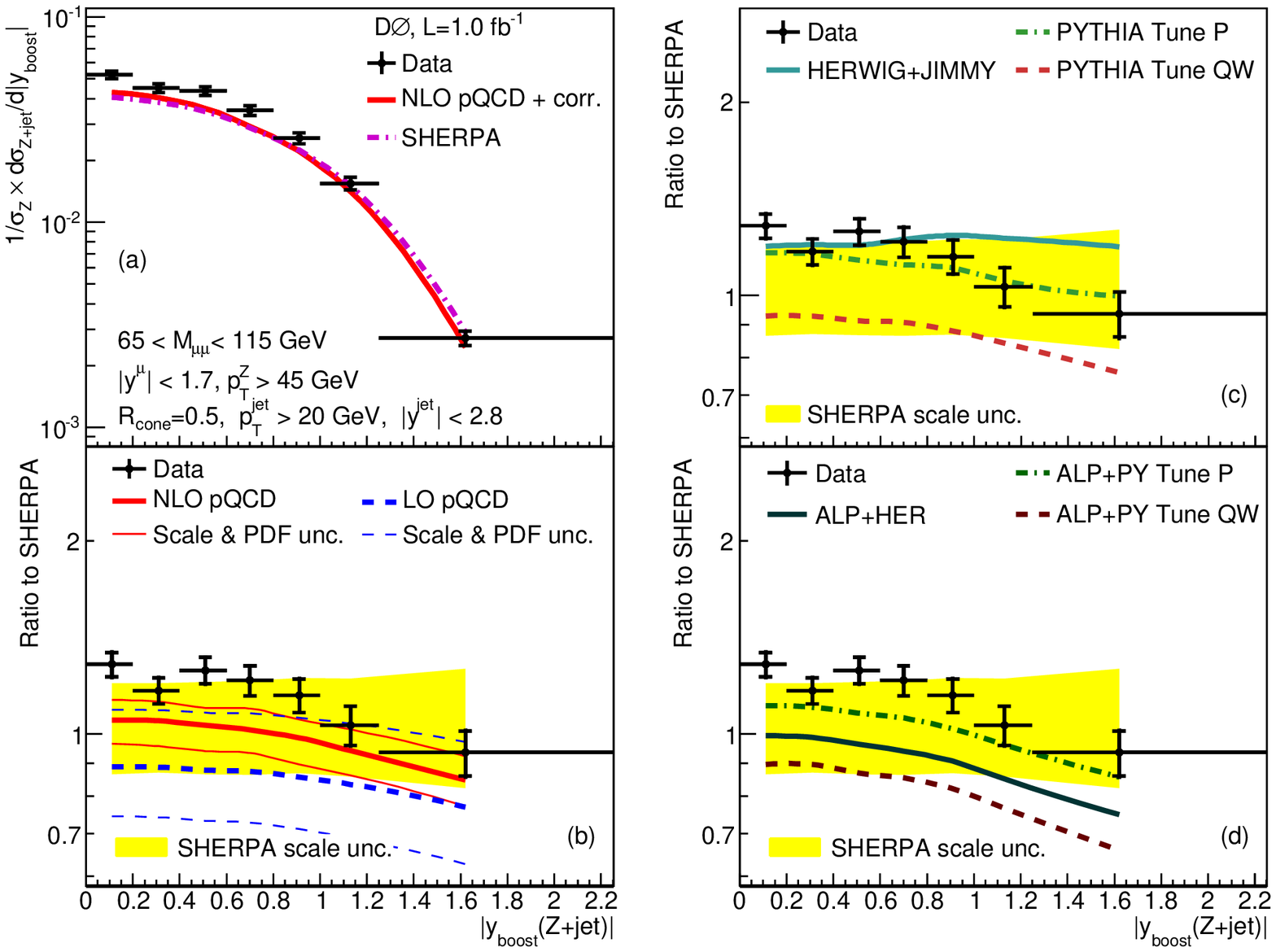}
\caption{The measured cross section in bins of $y_{boost}$ for \Zgammas + jet + X events with \pTZ larger than 45 \GeV.  Predictions from NLO pQCD and \textsc{SHERPA} are compared to the data in the  upper left plot.  The ratio of data and predictions from NLO pQCD, \textsc{ALPGEN}, \textsc{HERWIG} and \textsc{PYTHIA} to the prediction from \textsc{SHERPA} are shown in the lower plots.  } 
\label{fig13}
\end{figure}


\begin{thebibliography}{9}   % Use for  1-9  references
%\begin{thebibliography}{99} % Use for 10-99 references

\bibitem{Abazov:2007hk}
  V.~M.~Abazov {\it et al.}  [D0 Collaboration],
  %``A Combined search for the standard model Higgs boson at $\sqrt{s}$ =
  %1.96-TeV,''
  Phys.\ Lett.\  B {\bf 663}, 26 (2008)
  [arXiv:0712.0598 [hep-ex]].
  %%CITATION = PHLTA,B663,26;%%

\bibitem{Abazov:2009ii}
  V.~M.~Abazov {\it et al.}  [D0 Collaboration],
  %``Observation of Single Top-Quark Production,''
  Phys.\ Rev.\ Lett.\  {\bf 103}, 092001 (2009)
  [arXiv:0903.0850 [hep-ex]].
  %%CITATION = PRLTA,103,092001;%%

\bibitem{Campbell:2002tg}
  J.~M.~Campbell and R.~K.~Ellis,
  %``Next-to-leading order corrections to $W^+$ 2 jet and $Z^+$ 2 jet production
  %at hadron colliders,''
  Phys.\ Rev.\  D {\bf 65}, 113007 (2002)
  [arXiv:hep-ph/0202176].
  %%CITATION = PHRVA,D65,113007;%%


\bibitem{Mangano:2002ea}
  M.~L.~Mangano, M.~Moretti, F.~Piccinini, R.~Pittau and A.~D.~Polosa,
  %``ALPGEN, a generator for hard multiparton processes in hadronic
  %collisions,''
  JHEP {\bf 0307}, 001 (2003)
  [arXiv:hep-ph/0206293].
  %%CITATION = JHEPA,0307,001;%%

\bibitem{Gleisberg:2008ta}
  T.~Gleisberg, S.~Hoche, F.~Krauss, M.~Schonherr, S.~Schumann, F.~Siegert and J.~Winter,
  %``Event generation with SHERPA 1.1,''
  JHEP {\bf 0902}, 007 (2009)
  [arXiv:0811.4622 [hep-ph]].
  %%CITATION = JHEPA,0902,007;%%


\bibitem{Corcella:2000bw}
  G.~Corcella {\it et al.},
  %``HERWIG 6.5: an event generator for Hadron Emission Reactions With
  %Interfering Gluons (including supersymmetric processes),''
  JHEP {\bf 0101}, 010 (2001)
  [arXiv:hep-ph/0011363].
  %%CITATION = JHEPA,0101,010;%%

\bibitem{Sjostrand:2000wi}
  T.~Sjostrand, P.~Eden, C.~Friberg, L.~Lonnblad, G.~Miu, S.~Mrenna and E.~Norrbin,
  %``High-energy physics event generation with PYTHIA 6.1,''
  Comput.\ Phys.\ Commun.\  {\bf 135}, 238 (2001)
  [arXiv:hep-ph/0010017].
  %%CITATION = CPHCB,135,238;%%


\bibitem{Abazov:2008ez}
  V.~M.~Abazov {\it et al.}  [D0 Collaboration],
  %``Measurement of differential $Z / \gamma^{*}$ + jet + $X$ cross sections in
  %$p \bar{p}$ collisions at $\sqrt{s}$ = 1.96-TeV,''
  Phys.\ Lett.\  B {\bf 669}, 278 (2008)
  [arXiv:0808.1296 [hep-ex]].
  %%CITATION = PHLTA,B669,278;%%

\bibitem{Abazov:2009av}
  V.~M.~Abazov {\it et al.}  [D0 Collaboration],
  %``Measurements of differential cross sections of Z/gamma*+jets+X events in
  %proton anti-proton collisions at sqrt{s}=1.96 TeV,''
  Phys.\ Lett.\  B {\bf 678}, 45 (2009)
  [arXiv:0903.1748 [hep-ex]].
  %%CITATION = PHLTA,B678,45;%%

\bibitem{Abazov:2009pp}
  V.~M.~Abazov {\it et al.}  [D0 Collaboration],
  %``Measurement of Z/gamma*+jet+X$ angular distributions in ppbar collisions at
  %sqrt{s}=1.96 TeV,''
  arXiv:0907.4286 [hep-ex].
  %%CITATION = ARXIV:0907.4286;%%
 
\bibitem{Alwall}
  J.~Alwall {\it et al.},
  ``Comparative study of various algorithms for the merging of parton showers
  and matrix elements in hadronic collisions,''
  Eur.\ Phys.\ J.\  C {\bf 53}, 473 (2008)
  [arXiv:0706.2569 [hep-ph]].
  %%CITATION = EPHJA,C53,473;%%

\bibitem{Abazov:2005pn}
  V.~M.~Abazov {\it et al.}  [D0 Collaboration],
  %``The Upgraded D0 Detector,''
  Nucl.\ Instrum.\ Meth.\  A {\bf 565}, 463 (2006)
  [arXiv:physics/0507191].
  %%CITATION = NUIMA,A565,463;%%

\bibitem{Aaltonen:2007cp}
  T.~Aaltonen {\it et al.}  [CDF - Run II Collaboration],
  %``Measurement of Inclusive Jet Cross Sections in $Z/\gamma^* (\to ee)+$ jets
  %Production in $p\bar{p}$ Collisions at $\sqrt{s}=1.96$ TeV,''
  Phys.\ Rev.\ Lett.\  {\bf 100}, 102001 (2008)
  [arXiv:0711.3717 [hep-ex]].


\end{thebibliography}
\end{document}